# Comments on the report "Indications of anomalous heat energy production in a reactor device containing hydrogen loaded nickel powder." by G.Levi, E.Foschi, T.Hartman, B.Höistad, R.Pettersson, L.Tegnér, H.Essén.

Göran Ericsson[1], Stephan Pomp

*Division of applied nuclear physics, Uppsala University, Sweden*

**Abstract:** In a recent report titled "Indications of anomalous heat energy production in a reactor device containing hydrogen loaded nickel powder" and published on arXiv, G. Levi and co-workers put forth several claims concerning the operations and performance of the so-called E-Cat of inventor Andrea Rossi. We note first of all that the circumstances and people involved in the test make it far from being an independent one[2]. We examine the claims put forth by the authors and note that in many cases they are not supported by the facts[3] given in the report. We present results from thermal calculations showing that alternative explanations are possible were the authors seem to jump to conclusions fitting pre-conceived ideas[4]. In general we find that much attention is drawn to trivialities[5], while important pieces of information and investigation are lacking and seem not to have been conducted or considered. We also note that the proposed claims would require new physics in not only one but several areas. Besides a cold-fusion like process without production of any radiation also extreme new material properties would be needed to explain what rather seems to be a problem of correct measurement. Therefore, it is clear to us that a truly independent and scientific investigation of the so called E-Cat device, convincingly demonstrating an "anomalous heat energy production" has not been presented in the arXiv report[6] and is thus, to-date, still lacking.

### 1) Introduction and general comments

In a recent report titled "Indications of anomalous heat energy production in a reactor device containing hydrogen loaded nickel powder" and published on arXiv [1], G. Levi and collaborators put forth several claims concerning the performance of the so-called E-Cat of Andrea Rossi. Below we examine several of these claims. Our comments reported here are based on the report as uploaded on arXiv (http://arxiv.org/abs/1305.3913v2) and Elforsk home pages on May 20, 2013 (version 2) [1]. Any possible subsequent updates have not been considered.

The intended audience of the arXiv report is a bit unclear, but we must assume that the authors have intended to write a comprehensive technical report that contains all the relevant information and data in order to validate their observation of excess, anomalous heat. Judging by the names in the author list, they must certainly have been aware of the large degree of skepticism that surrounds Rossi and his claims. We would therefore have expected the authors to clearly describe not only how they have dealt with the technical details, but also any possibilities of fraud. However, as detailed below, we find the report lacking in both technical detail, experimental and analysis methods as well as in discussing sources of possible deception.

---

[1] E-mail: Göran.Ericsson@physics.uu.se
[2] Parts of the Abstract have been misunderstood as containing unmotivated strong criticism with no support in the main text we have collected a few examples to illustrate the validity/justifiability of the summary statements we use in our Abstract (and Conclusions) in an Appendix. See Appendix, Sect. A, for clarification concerning this specific statement.
[3] See Appendix, Sect. B.
[4] See Appendix, Sect. C.
[5] See Appendix, Sect. D.
[6] See Appendix, Sect. E.

Even though it is pitched as a technical (scientific) report, the report lacks some fundamental technical information regarding the three presented tests, such as:

- A description, motivation and discussion on the choice of measurement methods and procedures.
- Complete electrical wiring diagrams including the resistive heating coils, the control box and the measurement points, extending from input mains, and including ground and all three phases.
- Mechanical diagrams of the device, including geometry and material composition.
- Any data whatsoever from the "dummy" test mentioned on page 17 of Ref [1].
- We would also have expected a presentation of comprehensive tables giving all the validated, fundamental measurements done in the course of the tests – or possibly a reference to a web page where such information can be found and retrieved.

Since no access to the core of the device was granted, and thereby *no real scientific investigation* of the core processes were allowed, it is unclear why a group of scientists (including chemists, nuclear and theoretical physicists) should be assigned to perform this measurement. The task seems rather to require expertise in IR and electrical measurements. It would therefore seem more appropriate that an independent agency expert in such measurement techniques were assigned the task; in Sweden the Technical Research Institute, "Sveriges Tekniska Forskningsinstitut (SP)" could possibly be contacted. It would then also seem appropriate to use similar methods and measurement protocols as those in determination of the COP of heat pumps, where, as in the present case, the input is electrical power and the output is heat.

A major problem with this test is the many restrictions and conditions that seem to have been imposed by Rossi on the measurement group and their work. In our opinion, a truly independent test, even of a "black box" device, would mean work in our own laboratory, with our own equipment, with only written instructions (and possibly telephone support) by the "inventor", with a measurement method of our own choosing etc. The report indeed states that the aim of the work is "to make an independent test of the E-Cat HT reactor under controlled conditions ..." (pg 1). However, we note several circumstances that contradict this statement:

i. The first author, G. Levi, has been closely involved in numerous tests and promotions of the E-Cat together with the inventor, A. Rossi, over the last 2 ½ years. His independence is not as clear as one would wish,
ii. Several of the other authors, at least R. Pettersson and H. Essén, have also participated in previous demonstrations arranged by Rossi and have then to some degree committed to a positive appreciation of the device,
iii. The measurements were done in Rossi's premises in Italy with a corresponding loss of control over important factors in the measurement process,
iv. The "reactor" and its control circuits were operated by personnel assigned by Rossi (in the December test even started before the investigators were given access),
v. Measurements were done on (at least) two different types of devices,

vi. Estimating the heat output by a combination of IR camera measurements and convection calculations represents a new situation compared to previous tests, seemingly imposed by the circumstances (i.e. Rossi) rather than by choice.

With the "black box" approach imposed by Rossi (due to several "industrial trade secrets") in the present case, the measurements are restricted to total system energy (power) IN and energy OUT. Performing and reporting these measurements with the utmost care then becomes the one critical task in order to establish the existence, or not, of any anomalous heat production. In the present case, energy INPUT is supplied by the electrical resistor coils inside the reactor; in addition there is the alleged energy contribution by reactions in the nickel powder.

Regarding the INPUT electrical power we note:

- Considering the fundamental and crucial importance of the measurement of the input electrical power, it is rather surprising that the report is quite brief on the details of the electrical circuits and measurements. The lack of a clear circuit diagram has already been mentioned. Other concerns not discussed in the report are the possibility of DC power, the waveforms of voltage and current at various points in the system, the possibility of power through ground leads or other ways that undisclosed electrical power can be supplied to the device.
- Previous tests have reported important discrepancies between the electrical input power as claimed by Rossi and those actually measured by specialists with proper electrical measurement equipment, to the extent where no excess heat production could be inferred [2]. With the knowledge of such critical observations a much more thorough reporting on the electrical measurements should have been provided.
- To be more specific still, since the results of the expert measurements referred to in the previous paragraph seem to have deviated from what was claimed by Rossi by a factor of about 3, which happens to coincide with the excess heat observed also in the March test, we would have expected a clear description of how the risk of such inconsistencies was avoided, and even an involvement of the specialists from the SP institute.
- In view of these severe inconsistencies, the fact that the control unit providing the electrical power was "not available for inspection, inasmuch as they are part of the industrial trade secret" (pg 15) is even more disturbing.

The authors claim "the reaction is fueled by a mixture of nickel, hydrogen, and a catalyst, which is kept an industrial trade secret" (pg 1). In view of the secrets surrounding the reactor fuel powder we wonder:

- How can the authors know there is nickel inside the reactor?
- How can the authors know there is hydrogen inside the reactor?
- In addition, the reference to "industrial trade secrets" with regard to the composition of the "fuel" makes all speculation about what is powering the alleged reactions meaningless.

Energy OUTPUT is purely by heat – radiation and convection – since no radioactivity above background has been observed in any of the tests reported here.

- The report describes in some detail the IR camera measurements and convection calculations performed to estimate the energy OUT. At least for the March test, a number of additional validations of the IR measurements seem to have been performed: a contact measurement on the reactor surface by a thermocouple and the application of emissivity calibration pads. The absence of such important information, together with the factor 2 difference of the COP in the two measurements, makes the December test of less value in our opinion.
- To our understanding, the sensor of the IR camera actually provides an electrical signal proportional to the emitted power in its region of sensitivity. It would seem to us that this signal, in combination with the wavelength response of the sensor, should have been reported and used for the derivation of the total emitted power. It would seem that going via an inferred surface temperature of the emitting object is an unnecessary detour.
- Since the final energy density levels reported are much beyond any known chemical fuel and approaching or surpassing those of conventional nuclear fuel, we find the absence of any detailed reporting and discussions on (nuclear) radiation measurements a disturbing omission.

Regarding the need to provide an external electrical heat source we wonder:

- Why is such a heat source necessary? How is it controlled?
- Is there a threshold temperature for the internal "anomalous" process to start?
- If so, what is the observed value for this temperature and how is it recognized?
- In this context, data from the start-up phase of the device would be highly relevant. Why are such studies missing and in particular why are the observed surface temperatures for the start-up phase of the device not reported?

Below we look more closely on the various tests mentioned or described in [1]. The report mentions four such tests, two performed in Nov. and Dec. 2012 by Levi (and Foschi?), one test in March 2013 where all authors participated, and a dummy test.

2)  **Remarks on the November 2012 test**

In the November test, the reactor was destroyed but the authors still claim that "although the run was not successful [...] it demonstrated a huge production of excess heat, which could not be quantified" (pg 2).

- No data are presented to support the claim for "huge excess heat" in this test. How do the authors know there was any power output beyond what was supplied by the resistive heating?

The authors claim to know that the "fuel" powder in the November test was not evenly distributed laterally in the reactor, but rather concentrated at the two ends of the cylinder.

- In the December and March tests, the authors seem to have little or no control over the core of the reactor where the fuel is kept. So, how do they know this?
- Why was the test arranged in this way?
- Could the observed effects be explained by pure electrical heating?

In this test, the authors observed lighter and darker horizontal bands on the surface of the reactor during operation (Fig. 1, pg 2). They claim that five dark bands in the image "nicely match the areas overlying the resistor coils."

- How was the exact projection on the surface of the internal resistor coils determined, in order to make the association of dark bands with the resistor coils?
- Since there are 16 heating resistor coils it is not clear why there should be only five horizontal "footprints" seen on the surface.
- The authors make no attempt to investigate if instead the six (possibly seven) <u>bright</u> horizontal bands could in fact be the surface "footprint" of the resistor coils seen in this view. In any case, since on a cylindrical surface there must be an equal number of dark and bright bands, and unless the exact spatial location of the resistor coils is known in the plane projection view of the camera, there seems no compelling reason to assume that either dark or bright bands are caused by the coils.
- Since there are no thermal calculations reported, how do the authors know that the 1 kW (constant, presumably, since it is mentioned that the "pulsed" self-sustained mode was only active in the March test) was not sufficient to cause the observed effects of (light or dark) bands and melting?

In summary we note that the arguments and statements that Levi et al. put forth concerning the November 2012 test make it seem that the authors jump to conclusions that fit their preconceived idea[7], namely, that there is an anomalous heat production, the very question the authors have set out to investigate. We find this not to be a scientific approach to the problem.

3) **Remarks on the December 2012 test**

The emissivity of the surfaces measured by the IR camera was not known. The authors claim that the most conservative assumption is then to set an emissivity of 1, and offer an example to show so.

- To us, the fact that the emissivity of the studied surface was *not* established seems to be a fundamental lack of experimental procedure. It was shown in the March test that quite straight-forward measures can be taken in order to estimate the emissivity. The emissivity could easily have been obtained after the measurements, in a manner shown by the March tests. We can only wonder why such a fundamental, complementary measurement was not performed before the report was published.
- We can only speculate that one factor contributing to the different results obtained in the December and March tests might be the lack of knowledge of the emissivity in the December test.

4) **Remarks on the March 2013 test**

---

[7] Compare Appendix, Sect. C.

**Ericsson et al.,** *Comments on the report "Indications of anomalous heat energy production ..."*

The authors present a long discussion on the deviation of the time-dependence of the observed temperature/emitted power from that of a generic resistor (pgs 25-26). They further argue that this time-evolution is supporting evidence for the presence of an anomalous heat source in the E-Cat device. We do not find this discussion convincing.

- We have performed a thermal analysis using the physics simulation tool "COMSOL" for MatLab, using values for geometry, materials and electrical heating coils extracted from the report (March test, cylindrical part only[8]). In Fig. 1 we show results from a calculation including resistive heating only (810W for 2min, then 0W for 4 min). The shape of the temperature curve at the surface is quite similar to what is presented in Ref. 1, Plot 3. Since we have not included any anomalous heat source in our model, it seems clear that the observed shape of the temperature at the surface can simply be attributed to heat diffusion from the internal heat source(s) through the steel cylinder to the surface. As a matter of fact, there seems to be no reason whatsoever why the surface temperature curves should look anything like those of a generic resistor (as, e.g., claimed in the caption of Plot 3 on page 25 in Ref. [1]), when this resistor is buried deep within the device.
- Our modeling indicates that the observed time-dependence is perfectly consistent with a situation where the internal heat source is supplied purely by resistive heating. Thus, the argument that the *shape* of the temperature/power curves is an indication of an anomalous internal heat source is not correct. And even if this has no direct bearing on the existence (or not) of any anomalous heat source within the reactor, the lack of understanding of such a fundamental physical process does not inspire confidence in the presented results.
- It would seem to us that a careful thermal modeling and analysis should have been performed before far-reaching claims about the shape of the temperature/power curves were made. Furthermore, with sufficiently detailed thermal modeling, it could be possible to shed some light on the possible internal workings of the reactor, possibly serving as an independent corroboration of the energy balance calculations.
- From our thermal analysis it is also clear that if any form of pulsed, internal (electrical plus possibly "anomalous") power is supplied, the temperature (and hence the emitted power) at the surface of the reactor by necessity has to be out of phase with the power contributed by the internal sources. And obviously, the displayed pattern of surface heating and cooling clearly indicates that the internal heat source is not constant. However, the curves presented in [1] (especially Plot 8 on pg 27) show the electrical power and the surface emission power to be in phase. This can have two explanations. The simplest one is that the two curves have been forced "synchronized" for the sake of illustration by the authors. This should then have been clearly pointed out in the report, but, and this is more serious, it would then also constitute a deviation from reporting the actual experimental data as well as withholding important information that could contribute to an understanding of the device and its internal processes. The second possibility could be that the phase and amplitude of the additional anomalous heat source is such as to precisely synch the two curves. We leave it open if such a scenario can even be conceived, but it would of course be a rather remarkable coincidence, and could actually offer an independent possibility to assess some of the properties of the alleged anomalous source. At present we have not had the resources to take our modeling to that level.

---

[8] The COMSOL model can be supplied on request

- We note that the authors' reasoning in the case of the thermal signature shows a tendency to quickly jump to interpretations and conclusions that support the extra-ordinary claim, rather than to try to find more mundane explanations based on already known, standard physics. In this case, it seems that normal heat diffusion and transport is sufficient to explain the observation (except for the perfect synchronization of electrical power with surface emitted power). The observed thermal time-evolution gives no reason to resort to anomalous explanations.
- It is also remarkable that the authors seem not to reflect upon the need of a pulsed heat source. If a heat source is needed to start a process in a very small amount of "fuel" in the E-Cat, it seems relevant to reflect about how the process can be controlled and why it can be stopped by switching off the external heat source.

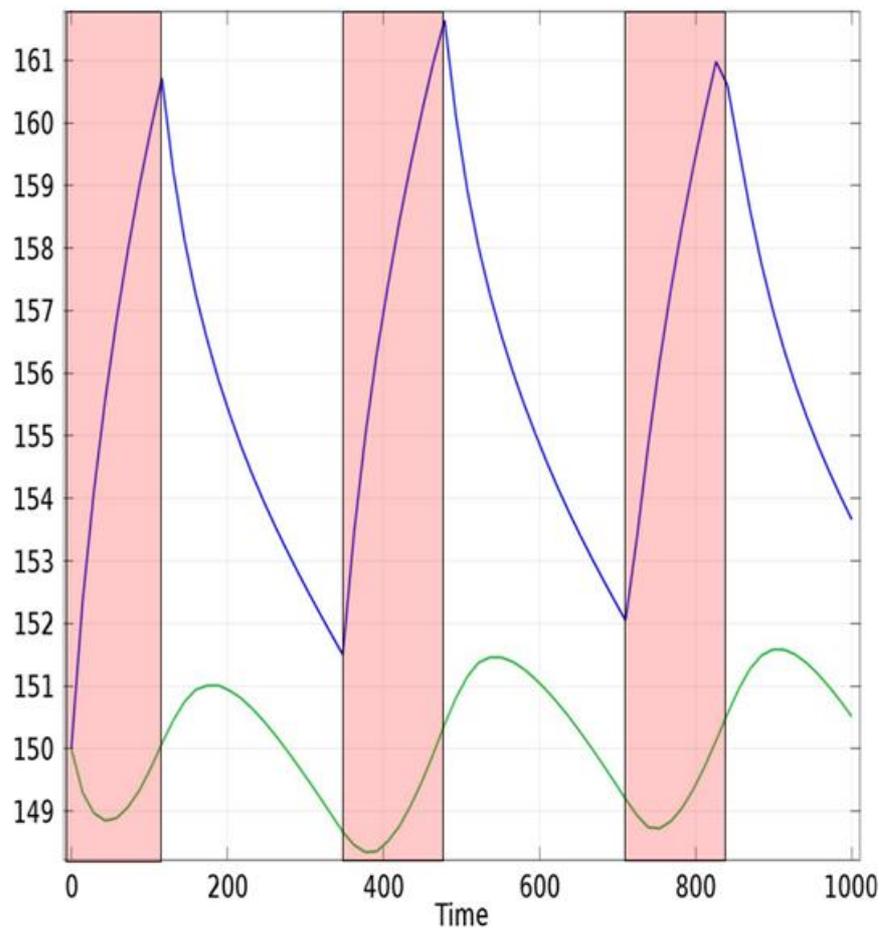

Fig. 1. Results from a COMSOL simulation of the heat transport in a E-Cat-like cylindrical device. Temperature in degrees Celsius (y axis), time in seconds (x axis). Resistive heating only at 810 W for 2 min (indicated by red bands), and then 4 min OFF. The blue (upper) curve gives temperature evolution of the heating region (resistor coils at R=15mm). The green (lower) curve shows the temperature evolution at the surface of the cylinder (at R=45mm).

We note that there are (at least) three wires going into the E-Cat in Fig. 10 of Ref. [1]. What is the function of each of them and how were they connected externally to the control box and internally in the device?

Data on electrical input power were provided from the continuous video recordings of the PCE-830 instrument. However, only an average value for the measurements during the resistive heating period is given. No data from the period with heating off are given.

- It is unclear to us why the operation of the device in this test is labeled "self-sustained" when it obviously needs a supply of external heating throughout the test. From our thermal analysis there seems to be no reason to assume that power is supplied outside of the 2 min electrical heating phase – although it does seem that more power than 810 W is required to obtain the observed surface temperature of about 300 degrees Celcius.
- We would have expected a diagram or table showing the power drawn from the grid over (at least) one full 6 minute cycle, including both the 2 minute phase of resistive heating and the 4 minute phase without any external power to the device.
- It would seem to us that the power consumed by the control box could be obtained from the power measurements during the phase when no external heating was supplied to the reactor device. Instead, the power to the control box was obtained from the dummy measurement. Why?
- We would have expected to see a discussion and data on power consumption all the way from the start-up phase of the device to the steady-state phase, in order to show how and when the "anomalous" heating source kicks in. After all, that source allegedly provides about twice as much power as the heating coils and we would have expected the signature of its onset to be presented and discussed.

The authors give their best estimate for the amount of fuel as 0.3 g. However, in most of their calculations they select not to use this value but rather an *ad hoc* value of 1 g. The reason for this seems to be to stay on the conservative side in their estimated energy density. However, we find this effort somewhat misguided as it is important to provide a best estimate of the effect with actual measured values; only after that one should proceed to discuss the validity of the measured values and assign appropriate uncertainties in order to estimate the range of possible values.

- Using the 0.3 g at face value, and assuming that the energy balance is correctly determined in the report, we find that the energy density of the fuel powder is about 250 kWh/g thermal. This can be compared with the thermal density of ordinary fission reactor fuel of about 750 kWh/g thermal. Thus, the fuel powder would under these assumptions give about 1/3 of the energy density of fission reactor fuel. Even with the ad hoc assumption of 1 g of fuel powder, the energy density is about 1/10 of nuclear reactor fuel, far beyond any known chemical compound.
- The authors half acknowledge the extremely high energy densities implied by their measurements, but do not carry the discussion to the logical end. The only processes we know of today that can give such energy densities are nuclear. This would normally be associated with strong emission of radiation, in particular gamma, but also neutrons, beta and even some heavier charges particles (depending on the exact nuclear transformations involved). It would also cause nuclear transmutations of the fuel. In view of this background, which must have been very clear to the authors as they include several experts in nuclear physics and measurements, it is surprising that the investigations of radiation emitted during the operation of the device are not presented as part of the report, but only referred to indirectly by a quote from a Dr Bianchini. In a truly scientific investigation, the analysis of the fuel would of course be the top priority – but unfortunately the restrictions imposed by Rossi on the present tests did not allow for this. It is unfortunate that the group of investigators agreed to such conditions during these tests. This is even more surprising as investigations of the fuel powder have been conducted and reported in the past, then with no indications of radioactivity or nuclear transmutations of the Nickel isotopes [3].

- Even more extreme seem to be the claimed power densities. If the fuel weighs 0.3 g and is supposed to produce on the order of 1667 W during the 96 hours of the December test (to get 160 net kWh, see pg 28), or 534 W during 116 hours of the March test (to get 62 net kWh, see pg 28), then power densities in the range of 1.78 to 5.56 MW/kg would have been achieved (the authors give a value of $7 \cdot 10^3$ W/kg on pg 28). That is far more than the power density obtained from fuel in today's light-water reactors (there is about 100 tons of fuel in a 3 GW (thermal) fission reactor, i.e. 0.03 MW/kg), and far beyond the y-axis scale in the power density plot shown as Fig 9 in Ref. [1]. The authors seem not to reflect on this remarkable claim.

### 5) Remarks on the "dummy" test

A "dummy" test was conducted in conjunction with the March test, after the measurements of the loaded reactor. This involved a reactor without powder charge and end caps. The test was performed with a continuous electrical input power of about 810 W.

- No data or figures associated with the "dummy" test are reported. This makes it very hard to assess the validity and usefulness of this measurement.
- The dummy test was not performed in the same way as the test of the "loaded" reactor and can therefore hardly be seen as a conclusive no-charge test. For example, in a proper dummy test, care should have been taken to supply input power in the same pulsed manner as in the loaded test. (It would then also have been clear to what extent the shape (time dependence) of the surface temperature curves as reported from the March test require any anomalous source, or can be explained as a simple consequence of heat diffusion.)

### 6) Conclusions and final comments

The authors of the report published on arXiv [1] claim to have performed an independent test of what is called an E-Cat HT reactor. We have shown that the tests lack many aspects of what would constitute an independent measurement, that the method chosen for testing seems unmotivated and, most of all, that the paper lacks several vital pieces of technical information that would be expected of such a report and, indeed, necessary to support the reported claims. We also find a few disturbing cases where the authors are jumping to the extra-ordinary conclusion instead of critically assessing other, more established physics explanations. We note that the proposed claims would require new physics in several areas. Besides a cold-fusion like process without production of any radiation also extreme new material properties would be needed to explain what rather seems to be a problem of correct measurement. We are surprised that the authors make such remarkable claims based on a report with so many shortcomings. We also find that much attention is drawn to trivialities while important pieces of information and investigation are lacking and seem not to have been conducted and considered. Wishful thinking seems to have replaced scientific rigor in some cases. Therefore, it seems clear to us that a truly independent and scientific investigation of the so called E-Cat HT device is still pending. Thus, we do not think that a convincing demonstration of "anomalous heat energy production"[9] has been presented in the arXiv report, or anywhere else to date.

**Acknowledgements**

Several people have contributed to the listed comments during discussions over coffee and via e-mail. We would like to thank especially Andrea Mattera for his help in providing the COMSOL simulations.

---

[9] See Appendix, Sect. E.

**APPENDIX**

Since parts of the Abstract of our "Comment" have been misunderstood as containing unmotivated strong criticism with no support in the main text we have collected a few examples to illustrate the validity/justifiability of the summary statements we use in our Abstract (and Conclusions). The examples are presented below under the heading of the sentence/statement in question.

**A) "Circumstances and people ... make it far from an independent one."**

We have elaborated this on pg 2 of our "Comment", in Points i) - vi).

**B) "Claims ... not supported by the facts given in the report."**

- Levi at al claim (Introduction, pg 1, 2nd paragraph): "*As in the original ECat, the reaction is fueled by a mixture of nickel, hydrogen, and a catalyst ...* " (Similar statements are also made in the Abstract.)

There are no facts in the report suggesting that the authors had any control over, had any access to or did any measurements of the composition of the "fuel" powder in the ECat. The statement regarding fuel composition is not based on facts given in the report, and no reference is given to back it up.

- Levi et al claim (Introduction, pg 2): "*... for the November test the charge in the inner cylinder was not evenly distributed but concentrated in two distinct locations along the central axis.*"

There is no description, motivation or discussion of the handling of the powder and its application into the device to accomplish this "uneven" distribution – or why such an arrangement was chosen. As a matter of fact, in the December and March tests the authors seem to have had no access to or control over the powder. It seems the uneven distribution of the powder is not a fact established by the authors.

- Levi et al. claim (Introduction, pg 2): "*Although the run was not successful as far as obtaining complete data ... it demonstrated a huge production of excess heat...*"

This is a purely anecdotal statement, with no facts to back it up. This is further problematic since, already in the Introduction, it seems to establish as a fact the phenomenon the report sets out to investigate.

- Levi et al. claim (Introduction, Figs 1-2): "*The darker bands in the photograph are actually the shadows of the resistor coils.*"

This claim, which is used as corroborating evidence for the anomalous heat production, is given without any supporting analysis, motivation or explanation. (See further below.)

**C) "Jumping to conclusions fitting pre-conceived ideas"**

One serious problem with the Levi paper is the absence of a null hypothesis – or ANY alternative hypothesis for that matter. When investigating a new, controversial phenomenon, the obvious null hypothesis is that there is NO anomalous effect. An attempt should then be made to explain the observations of the investigation with this as a starting point. The Levi report only presents interpretations of their observations that favour the claim of anomalous heat production and makes no attempt to contrast this with a null hypothesis (of any kind). Four examples of this are presented below.

- Levi et al. conclude: (Introduction, pg 3, 2nd paragraph). "... *the temperature dips visible in the diagram are actually shadows of the resistor coils projected on the camera lens by a source of energy located further inside the device* ..."

This claim is not supported by any analysis, motivation or explanation. The claim is presented as corroborating evidence for the "anomalous heat" effect (although presented already in the Introduction). As such we would expect it to be presented with facts, discussing coil positioning in the device, device alignment in the measurement rig, calculations of the projection of internal coils onto the surface and finally the projection into the view of the IR camera – or as a minimum a statement that such an investigation has been conducted. Furthermore, since the idea that a metal object inside an insulator in a heat diffusion context would cause this kind or pattern on the surface is by no means trivial, a thermal transport calculation or similar argument should have been presented to make it plausible. However, no facts, calculations or arguments are presented. The only evidence presented is the distance between "dips", which as we point out, is equally true for the interspacings between the coils as for the coils themselves. It seems clear that without strong supporting arguments along the lines outlined here, an explanation based on the null hypothesis, namely, that the observed bright bands are produced by the heating effect of the energized coils, is at least as plausible as the one proposed by the authors. But the authors put forward only the explanation that involves an internal anomalous heat source.

- Levi et al. conclude (pg 25, last paragraph): "*If we compare [*Plot 3*] in detail with the standard curves of a generic resistor (Plot 4 and Plot 5), we see that the former differ from the latter in that they are not of the exponential type. What appears obvious here is that the priming mechanism pertaining to some sort of reaction inside the device speeds up the rise in temperature, and keeps the temperature higher during the cooling phase*."

This interpretation, that the shape of the temperature curves is corroborating evidence for the anomalous heating effect, is presented without any supporting thermal analysis pertaining to the device under investigation. We have performed and presented such thermal transport calculations showing that the shape of the temperature curve can be explained by the null hypothesis, namely that heating is provided ONLY by the resistor coils during their ON cycle. In order to explain not only the shape of the temperature curve but also its absolute value, it seems that the power during the resistor ON period needs to be about 3 times higher than reported. In view of the report's quite brief description of the electrical system and measurements, a possibility could be that undisclosed electrical power was supplied to the ECat resistor coils in a manner not detected by the test team. The authors seem not to consider these possibilities.

- Levi et al. remark (pg 25, 4th paragraph): "*An interesting aspect of the ECat HT2 is certainly its capacity to operate in self-sustaining mode*."

This statement is given as if the operation in a "Self-sustained" mode had been established. We emphasize again that NO facts about the ECat have to date been established, so all claims about the operational modes of this device have to be backed up by facts. And as discussed above, the shape of the temperature curve can be explained by heating during the electrical coil ON period only. Our thermal calculations also indicate that both the shape and the absolute level can be explained if all the power is supplied during the 2 minute ON phase. Again there are alternative explanations for what is observed which do not involve any "anomalous heat source". The authors seem not to consider these possibilities. (As an aside, we must also question the authors use of the expression self-sustained, when the device obviously needs to be supplied with external power at a 33% duty cycle all through its 100 hours operation.)

- In addition, some statements in the Introduction chapter of the Levi report suggest that the authors accept as real the phenomenon they are set out to investigate. And let us be very clear: the anomalous heat production of the ECat device is NOT an established scientific/technical fact. The Levi report is the first attempt to investigate this phenomenon in a scientific manner. It is then problematic if the authors already have a clear idea of what they are about to find in the course of their investigation. For example, Levi at al claim (Introduction, 1st paragraph): "*When the container was heated, substantial heat was produced in excess of the input heat.*" and (Introduction, 2nd paragraph): "*The charge sets off the production of thermal energy after having been activated by heat produced by a set of resistor coils ... Once operating temperature is reached, it is possible to control the reaction by regulating the power to the coils.*" These statements are given without reference or qualifications so would seem to be statements the authors hold for true.

D) **"… much attention is drawn to trivialities while important pieces of information and investigation are lacking …"**

We have listed several important pieces of information that are missing from the report, such as an electrical wiring scheme, mechanical diagrams of the device, time resolved raw data tables, motivations for several statements and claims etc. In contrast, the 28 page report contains 10 photographs, some of which seem to convey essentially the same information (Figs 4, 6, 11), a set of three tables (Table 1-3) containing the same information, several Tables copied from reference sources (Tables 4-5), two Ragone charts copied from References, none of which contain the results obtained in this test (Figs 9, 15) and plots of unclear value (Plot 9).

E) **A final remark regarding "Conclusions".**

Throughout the Levi report the authors quote measured values with 10-15% uncertainty, presumably mostly of systematic origin. To us, this means that the authors considered their control over the experimental situation good.

Levi et al. conclude (Conclusions, pg 28): "*Both [tests] have indicated heat production from an unknown reaction primed by heat from resistor coils. The results obtained indicate that energy was produced in decidedly higher quantities than what may be gained from any conventional source.*"

Here, as well as in the title, the use of the word "indicate" seems to be a way for the authors to convey an uncertainty in the presented results that goes beyond their stated error margins. If so, however, this is never further elaborated upon in the text. Rather to the contrary, the authors continue their conclusions:

"[…] *the results obtained place both devices several orders of magnitude outside the bounds of the Ragone plot region for chemical sources.*" … "*Even from […] the most conservative figures on energy production, we still get a value […] that is one order of magnitude higher than any conventional source.*"  … "*[…] the energy densities that were measured should be considered as lower limits of real values.*"

In view of the reported 10-15% uncertainty levels and the conclusions cited above, it would seem that the authors have clearly established the effect of "anomalous heat production". Therefore we find the use of the word "indications" problematic, since it seems to be at odds with what is actually reported and concluded. It is unclear to us why such a label should be used in a situation as described in the Levi report. We think it is important to state conclusions (and title) in a way consistent with the contents of the report, and we see no facts presented to warrant the use of "indications". If there are further sources of uncertainty known to the authors, underlying this choice, they should be clearly presented in the report. And if the authors suspect even more serious problems, such as undisclosed sources of power, it is our opinion that they should have refrained from publishing until such problems are duly investigated and cleared.